\documentclass[twocolumn,floats,showpacs,superscriptaddress,pre]{revtex4}
\usepackage{amsmath,amssymb,bm}
\usepackage{graphics}
\usepackage{epsfig}
\usepackage{graphicx}

\begin{document}

\title{On the Quantum Creation of Matter in the Expanding Universe}
\author{Natalia Gorobey and Alexander Lukyanenko}
\email{alex.lukyan@rambler.ru}
\affiliation{Department of Experimental Physics, St. Petersburg State Polytechnical
University, Polytekhnicheskaya 29, 195251, St. Petersburg, Russia}

\begin{abstract}
Quantum Action Principle which has been used as a ground for a probabilistic
interpretation of one-particle relativistic quantum mechanics \cite{GLL} is
applied to quantum cosmology. The quantum creation of matter in a
minisuperspace model with one homogeneous scalar field is considered. The
initial state of the universe is defined as a stationary ground state of the
Hamiltonian with the Euclidean signature in which the mean value of the
universe radius is equal to the Plank length and the number of the scalar
field quanta is equal zero. We interpret the change of the signature as the
universe "birth". From this moment of time the dynamics of the scale factor
is considered as classical. The real phase of the amplitude of the creation
process is taken as a quantum action. The balance between matter and
gravitation energies in the creation process is fulfilled by the condition
of the stationarity of the quantum action with respect to the internal time
of the universe.
\end{abstract}

\maketitle
\date{\today }





\section{\textbf{INTRODUCTION}}

The idea of the quantum creation of matter in the expanding universe has
been developed in 1970's \cite{GMM}. The creation was considered as a
parametric excitation of quantum matter fields on the background of the
classical Friedman expanding universe. But the effectiveness of this
mechanism of excitation is not high \cite{W}, at least for the ordinary
matter. The rate of the matter creation is maximum near the singularity, but
the latter is a problem in classical cosmology. This approach by itself is
considered as an approach to precise quantum cosmology (QC) with the
quantized space-time metrics. In the precise theory constraints (quantum
constraints) play the central role and determine the dynamics of the
universe \cite{MTW}. Specifically, the Hamiltonian constraint ensures the
balance between matter and gravitational field energies in the universe.
This balance was not taken into account in the approach mentioned above \cite%
{GMM}. To take it into account we don't need precise QC. The goal may be
achieved by means of a quantum action principle (QAP) in which an analog of
the classical constraints is obtained at the quantum level as a set of
conditions of stationarity of a quantum action \cite{GL}. From the first
time QAP was used for a probabilistic interpretation of one-particle
relativistic quantum mechanics in \cite{GLL}. A quantum action may be
builded in a semi-classical form with a classical gravitational part and a
quantum matter addition. This semi-classical quantum action will define the
dynamics of the classical geometry of the universe with account of a quantum
matter back-reaction, and it has to be stationary with respect to Lagrangian
multipliers $N_{\mu }$, which are coefficients in front of the classical
constraints in the classical action of General Relativity \cite{MTW}.
However, the situation near the singularity must be considered in the
framework of precise QC. The present work is devoted to application of QAP
to the simplest minisuperspace model of the universe - the Friedman universe
with one homogeneous scalar field. We make two improvements in the original
approach to the quantum creation of matter. The first one - we consider the
scalar field as the only matter contents of the universe, and take into
account a back-reaction of its quantum dynamics on the classical dynamics of
the scale factor. The second - the energy balance between the matter and
geometry is ensured by a condition of stationarity of a quantum action with
respect to an internal proper time of the universe. In order to solve the
problem of the initial state of the universe near the singularity, in the
next section we begin with formulation of full quantum theory of the
Friedman universe. Then we come to a semi-classical description of the
dynamics of the Friedman universe in analogy with the dynamics of the
Minkowsky time parameter $x^{0}$ of a relativistic particle, or a bosonic
string \cite{GL}.

\section{MINISUPERSPACE MODEL OF THE UNIVERSE}

The classical dynamics of the Friedman universe with the space-time interval
\begin{equation}
ds^{2}=N^{2}\left( t\right) dt^{2}-a^{2}\left( t\right) \left[ d\chi
^{2}+\sin ^{2}\chi \left( d\theta ^{2}+\sin ^{2}\theta d\varphi ^{2}\right) %
\right] ,  \label{1}
\end{equation}%
where $N\left( t\right) $ is the lapse function, and $a\left( t\right) $ is
the spatial scale factor, with one scalar field $\phi $ is described by the
classical action (velocity of light equals unity) \cite{MTW}: 
\begin{equation}
I=\int \left[ \frac{1}{2g}\left( aN-a\frac{\overset{\cdot }{a}^{2}}{N}%
\right) +\frac{1}{2}2\pi ^{2}a^{3}\left( \frac{\overset{\cdot }{\phi }^{2}}{N%
}-m^{2}N\phi ^{2}\right) \right] dt,  \label{2}
\end{equation}%
where $g\equiv 3\pi /2G,$ $G$ is the Newton gravitational constant. The
canonical form of the action (\ref{2}) is 
\begin{equation}
I=\int\limits_{0}^{C}dc\left( p_{a}\overset{\cdot }{a}+p_{\phi }\overset{%
\cdot }{\phi }-H\right) ,  \label{3}
\end{equation}%
where 
\begin{equation}
H\equiv -\frac{1}{2}\left( \frac{gp_{a}^{2}}{a}+\frac{a}{g}\right) +\frac{1}{%
2}\left( \frac{p_{\phi }^{2}}{2\pi ^{2}a^{3}}+2\pi ^{2}a^{3}m^{2}\phi
^{2}\right)  \label{4}
\end{equation}%
is the Hamiltonian constraint, which in fact regulates the balance of matter
and gravitational field energies at the classical level. It must be equal
zero as a condition of stationarity with respect to the Lagrangian
multiplier $N$ \ \cite{MTW}. In equation (\ref{4}) we introduced a time
parameter $c$, $c\in \left[ 0,C\right] $, which is related to the ordinary
time as follows, $dc=Ndt$ \cite{Fo}. In this case the upper limit is a free
dynamical variable. 

Let us turn to quantum theory. Introducing the operators of momenta, 
\begin{equation}
\widehat{p}_{a}\equiv \frac{\hbar }{i}\frac{\partial }{\partial a},\widehat{p%
}_{\phi }\equiv \frac{\hbar }{i}\frac{\partial }{\partial \phi },  \label{5}
\end{equation}%
we define the Hamiltonian operator as follows: 
\begin{equation}
\widehat{H}=-\widehat{H}_{a}+\widehat{H}_{\phi }  \label{6}
\end{equation}
\begin{equation}
\widehat{H}_{a}\equiv \frac{1}{2}\left( -\hbar ^{2}g\frac{1}{a^{q}}\frac{%
\partial }{\partial a}a^{q-1}\frac{\partial }{\partial a}+\frac{a}{g}\right)
\label{7}
\end{equation}
\begin{equation}
\widehat{H}_{\phi }\equiv \frac{1}{2}\left( -\frac{\hbar ^{2}}{2\pi ^{2}a^{3}%
}\frac{\partial ^{2}}{\partial \phi ^{2}}+2\pi ^{2}a^{3}m^{2}\phi ^{2}\right)
\label{8}
\end{equation}%
Here we chose a certain ordering of non-commuting operator multipliers in (%
\ref{7}), assuming the integration measure on the minisuperspace $\left(
a,\phi \right) $ to be 
\begin{equation}
a^{q}dad\phi .  \label{9}
\end{equation}%
We interpret the scale factor $a,a\in \left[ 0,\infty \right) $ as a
"radial" coordinate in the superspace, and take for definiteness $q=3$.
Quantum dynamics of the universe in our theory is described by a wave
function $\psi \left( c,a,\phi \right) $, which \ is a solution of the Schr%
\"{o}dinger equation 
\begin{equation}
i\hbar \frac{\partial \psi }{\partial c}=\widehat{H}\psi .  \label{10}
\end{equation}%
with corresponding initial data. Let us mentione that $c\in \left[ 0,C\right]
$, and $C$ is up to now arbitrary.

The dynamical parameter $C$ is not observable and must be excluded in the
framework of QAP \cite{GL}. The role of observable time in our theory will
play the scale factor $a$. But one can use the description of the quantum
dynamics in terms of the formal time parameter $c$ in the Schr\"{o}dinger
equation (\ref{10}) for regularization of the theory near the singularity $%
a=0$. In order to make this regularization, let us slightly "improve" the
theory by introducing a signature function in the Hamiltonian (\ref{6}): 
\begin{equation}
\widehat{H}\left( c\right) =f\left( c\right) \widehat{H}_{a}+\widehat{H}%
_{\phi },  \label{11}
\end{equation}%
where, for example, 
\begin{equation}
f\left( c\right) =2\exp \left( -\frac{c}{c_{pl}}\right) -1.  \label{12}
\end{equation}%
Near the singularity, $c\rightarrow 0$, the "improved" Hamiltonian (\ref{11}%
) is a positive definite operator. It will be used in the next section for
determining the initial quantum state of the universe at the moment $c=0$.
After the Plankian epoc, $c>>c_{pl}$, the Hamiltonian (\ref{11}) restores
its Lorentzian signature.

\section{INITIAL QUANTUM STATE OF THE UNIVERSE}

Let the initial state of the universe $\psi _{0}\left( a,\phi \right) $ be
an eigenfunction of the Hamiltonian (\ref{11}) taken at the moment $c=0$:
\begin{equation}
\widehat{H}\left( 0\right) \psi _{0}=E\psi _{0}  \label{13}
\end{equation}%
which corresponds to its minimal eigenvalue $E$. In analogy with ordinary
quantum theory of the hydrogen atom, we expect that the ground state $\psi
_{0}\left( a,\phi \right) $\ of the universe is non-singular. In order to
find the ground state one can use the ordinary variational principle for the
functional 
\begin{equation}
F\left[ \psi \right] \equiv \frac{\left( \psi ,\widehat{H}\left( 0\right)
\psi \right) }{\left( \psi ,\psi \right) }.  \label{14}
\end{equation}%
An approximation to the exact ground state is given by the following
two-parametric Gauss function: 
\begin{equation}
\widetilde{\psi }_{0}\left( a,\phi \right) =\exp \left( -\frac{\alpha a^{2}}{%
2}-\frac{\beta \phi ^{2}}{2}\right) ,  \label{15}
\end{equation}%
for which the functional (\ref{14}) equals 
\begin{eqnarray}
F &=&\frac{3}{8g}\sqrt{\frac{\pi }{\alpha }}\left( \hbar ^{2}g^{2}\alpha
^{2}+1\right)   \label{16} \\
&&+\frac{\hbar ^{2}}{8\pi \sqrt{\pi }}\alpha ^{3/2}\beta +\frac{15}{16}\frac{%
\pi ^{5/2}m^{2}}{\alpha ^{3/2}\beta }.  \notag
\end{eqnarray}%
Its minimum value equals 
\begin{equation}
F_{m}=\frac{3^{1/4}}{2}\sqrt{\pi }\sqrt{\frac{\hbar }{g}}+\frac{\sqrt{15\pi }%
}{4\sqrt{2}}\hbar m,  \label{17}
\end{equation}%
when 
\begin{equation}
\alpha =\frac{1}{\sqrt{3}\hbar g},\beta =\sqrt{\frac{15}{2}}3^{3/4}\pi
^{2}mg^{3/2}\sqrt{\hbar }  \label{18}
\end{equation}%
The main result of our consideration is the estimation of mean value of the
scale factor in the initial ground state of the universe: it has a non-zero
value of the order of the Plank length, 
\begin{equation}
\left\langle a\right\rangle \equiv \frac{\left( \psi _{0},a\psi _{0}\right)
}{\left( \psi _{0},\psi _{0}\right) }=\frac{3^{5/4}}{4}\sqrt{\pi }\sqrt{%
\hbar g}\simeq l_{pl}.  \label{19}
\end{equation}%
The energy of the scalar field in the ground state (the second term in (\ref%
{17})) is close to the vacuum energy $\left( 1/2\right) \hbar m$. Therefore,
in analogy with the ordinary quantum theory of the hydrogen atom the initial
ground state of the universe is non-singular. The universe will remain in
the ground state up to the moment $c=$ $c_{pl}$, when the signature of the
Hamiltonian (\ref{11}) wiill be changed. This moment may be interpreted as
"birth" of the universe. Now one can go to the semi-classical description of
the universe dynamics.

\section{CREATION OF MATTER IN THE EXPANDING UNIVERSE}

Let us turn to the second stage of the universe dynamics after its "birth".
Let us shift the initial moment of the proper time to the moment of "birth",
and consider the universe dynamics on the interval $c\in \left[ 0,C\right] $
in the semi-classical approximation proposed in \cite{GL}. The
semi-classical approximation in the framework of QAP is achieved by means of
an "improvement" of the original classical action before quantization. We
modify the first part (\ref{7}) of the Hamiltonian, introducing additional
complex variables $\lambda =\lambda _{1}+i\lambda _{2}$ and $d=d_{1}+id_{2}$
for description of the scale factor dynamics, as follows: 
\begin{equation}
\widehat{H}_{a}^{\prime }=-\frac{1}{2g}\left\vert d\right\vert ^{2}+\lambda
\left( d-gp_{a}-ia\right) +\overline{\lambda }\left( \overline{d}%
-gp_{a}+ia\right) .  \label{20}
\end{equation}%
It is obvious that at the classical level the modified Hamiltonian is
equivalent to the original one, if we consider $\lambda ,d$ as independent
dynamical variables with canonical momenta equal zero. But now the
Hamiltonian becomes linear with respect to $p_{a}$. The idea \cite{GL} is to
quantize the theory with the arbitrary variables $\lambda \left( c\right)
,d\left( c\right) $ at the condition that they are fixed at the quantum
level in the framework of QAP. The corresponding modified Schr\"{o}dinger
equation may be written in a form: 
\begin{eqnarray}
&&\left( i\hbar \frac{\partial }{\partial c}-2i\hbar g\lambda _{1}\frac{%
\partial }{\partial a}\right) \psi  \label{21} \\
&=&\left[ -\frac{1}{2ga}\left( d_{1}^{2}+d_{2}^{2}\right) +\left( \lambda
_{1}d_{1}-\lambda _{2}d_{2}\right) +2a\lambda _{2}+\widehat{H}_{\phi }\right]
\psi .  \notag
\end{eqnarray}%
Now, one can consider the scale factor as a function of time $a\left(
c\right) $ and the expression in the round brackets in the left hand side of
(\ref{21}) as a full derivative of a wave function $\psi \left( c,a\left(
c\right) ,\phi \right) $ with respect to $c$, if we take $-2g\lambda
_{1}\equiv \overset{\cdot }{a}$. The function $a\left( c\right) $ must be
defined by means of QAP, as well.

The quantum action in QAP is defined as the real phase of the transition
amplitude for a given quantum transition \cite{GL}. Let us consider the
transition of the quantum oscillator $\phi $ from the (normalized) vacuum
state $\left\vert 0\right\rangle $, 
\begin{equation}
\left\vert 0\right\rangle =\pi ^{-1/4}\exp \left( -\frac{1}{2}2\pi
^{2}a_{0}^{3}\frac{m}{\hbar }\phi ^{2}\right) ,  \label{22}
\end{equation}%
at the moment $c=0$ with $a_{0}=l_{pl}$ to an excited (normalized) state $%
\left\vert n\right\rangle $ at the moment $c=C$, 
\begin{equation}
\left\vert n\right\rangle =\frac{\pi ^{-1/4}}{\sqrt{2^{n}n!}}H_{n}\left(
\sqrt{2\pi ^{2}a_{1}^{3}\frac{m}{\hbar }}\phi \right) \exp \left( -\frac{1}{2%
}2\pi ^{2}a_{1}^{3}\frac{m}{\hbar }\phi ^{2}\right) ,  \label{23}
\end{equation}%
where $a_{1}\equiv a\left( C\right) $ is a final value of the scale factor.
It is this quantity that will play the role of an observable time parameter
in the expanding universe. The corresponding transition amplitude 
\begin{equation}
A_{n0}\equiv \left\langle n\right\vert \widehat{U}_{C}\left\vert
0\right\rangle ,  \label{24}
\end{equation}
where $\widehat{U}_{C}$, is the evolution operator of the modified Schr\"{o}%
dinger equation (\ref{21}) in the interval $\left[ 0,C\right] $, can be
written in a form: 
\begin{equation}
A_{n0}=\exp \left( \frac{i}{\hbar }\Lambda _{Ca}\right) \left\langle
n\right\vert \widehat{U}_{C\phi }\left\vert 0\right\rangle ,  \label{25}
\end{equation}%
where $\widehat{U}_{C\phi }$ is the evolution operator of the Schr\"{o}%
dinger equation for only scalar field part on a classical homogeneous
space-time background with arbitrary function $a\left( c\right) $: 
\begin{equation}
i\hbar \frac{\partial \psi }{\partial c}=\widehat{H}_{\phi }\psi ,
\label{26}
\end{equation}%
and 
\begin{equation}
\Lambda _{Ca}\equiv \int\limits_{0}^{C}dc\left[ \frac{1}{2ga}\left(
d_{1}^{2}+d_{2}^{2}\right) +\left( \frac{\overset{\cdot }{a}}{2g}%
d_{1}+\lambda _{2}d_{2}\right) -2a\lambda _{2}\right] .  \label{27}
\end{equation}%
At this stage we can find the stationary value of the phase (\ref{27}) as a
function of additional variables $d_{1,2},\lambda _{2}$. A resulting
quantity is equal to the gravitational part of the original classical action
(\ref{2}).

It is the equation (\ref{26}), that describes the creation of matter in the
expanding universe (minisuperspace model) in the approach mentioned above
\cite{GMM}, at the condition that the dynamics of $a\left( c\right) $ is
classical. The parametric excitation of the scalar field $\phi $ arises due
to the dependence of $\widehat{H}_{\phi }$ on $a\left( c\right) $ in
accordance with (\ref{8}). If we only take into account this mechanism of
excitation, we will lose the balance between matter and gravitation field
energies, that is regulated by the equation (\ref{4}) in classical theory.
We shall restore this balance by taking into account the dynamics of the
scale factor $a\left( c\right) $ in the framework of QAP, and the important
additional condition of stationarity of a quantum action with respect to $C$%
. Therefore, in our approach  we have overcame only half of the way.

Let us turn to the formulation of QAP. We define the full quantum action as
a sum, 
\begin{equation}
\Lambda =\Lambda _{Ca}+\Lambda _{C\phi } .  \label{28}
\end{equation}%
Here $\Lambda _{Ca}$ is the classical action for the scale factor $a\left(
c\right) $, $\Lambda _{C\phi }$ is the real phase of the transition
amplitude $\left\langle n\right\vert \widehat{U}_{C\phi }\left\vert
0\right\rangle $ written in the exponential form: 
\begin{equation}
\left\langle n\right\vert \widehat{U}_{C\phi }\left\vert 0\right\rangle
=R_{C\phi }\exp \left( \frac{i}{\hbar }\Lambda _{C\phi }\right) .  \label{29}
\end{equation}%
In analogy with classical action principle, we formulate QAP as a set of the
stationarity conditions of the quantum action: 
\begin{equation}
\frac{\delta \Lambda }{\delta a\left( c\right) }=0,\frac{\partial \Lambda }{%
\partial C}=0.  \label{30}
\end{equation}%
The first equation in (\ref{30}) defines the (semi-classical) dynamics of
the scale factor $a\left( c\right) $ with fixed boundary values $a\left(
0\right) =a_{0},a\left( C\right) =a_{1}$. This equation of motion takes into
account the back-reaction of quantum dynamics of the scalar fied $\phi $.
The second equation in (\ref{30}) fixes the proper time $C$. It is this
condition, that restores the balance of energies in the creation of matter
process. The balance is restored by tuning of the proper time $C$ to given
parameters of the final state of the universe, i.e. the quantum number $n$,
and the spatial size of the universe $a_{1}$. Let us stress that the
semi-classical history of the scale factor and the length of the proper time
interval are defined for a given quantum transition $\left\vert
0\right\rangle \rightarrow \left\vert n\right\rangle $. We must substitute
these quantities into the amplitude (\ref{29}).

Calculation of probabilities in QAP needs a special consideration. The
quantum evolution of matter described by the Schr\"{o}dinger equation (\ref%
{26}) is unitary. But additional conditions (\ref{30}) and a posteriory
manipulations with amplitude (\ref{29}) destroy the unutarity. We must
renormalize all amplitudes (\ref{29}) ($n=0,1,2,...$) after taking into
account the back-reaction of the creation of matter process on the dynamics
of the scale factor and the energy balance. The properly normalized
transition amplitude of the process $\left\vert 0\right\rangle \rightarrow
\left\vert n\right\rangle $ is 
\begin{equation}
K_{n}\left( a_{1}\right) =\frac{1}{\sqrt{Z}}\left\langle n\right\vert
\widehat{U}_{C\phi }\left\vert 0\right\rangle ,Z\equiv
\sum\limits_{n=0}^{\infty }\left\vert \left\langle n\right\vert \widehat{U}%
_{C\phi }\left\vert 0\right\rangle \right\vert ^{2}.  \label{31}
\end{equation}%
This is the probability density to detect $n$ quanta of the homogeneous
scalar field $\phi $ at the moment when the radius of the universe will be
equal $a_{1}$. We expect that the probability will be maximum for states
with energy equal to the gravitational energy corresponding to a macroscopic
value of the scale factor $a_{1}$.

\section{CONCLUSIONS}

\label{c}

The balance between matter and\textbf{\ }gravitational energies in the
universe, which is guaranteed by the Hamiltonian constraint equation in
classical\ General Relativity is realized in the present work at the quantum
level as a condition of the stationarity of a quantum action with respect to
an internal time of the universe. The latter is not observable quantity in a
quantum universe. The role of observable time in the universe plays its
spatial scale factor $a$. We expect that the properly normalized transition
amplitude (\ref{31}), which takes into account the energy balance, will give
us a sufficient rate for the quantum creation of matter in the expanding
universe.

\begin{acknowledgments}
We are thanks V. A. Franke and A. V. Goltsev for useful discussions.
\end{acknowledgments}




\end{document}